\documentclass{vow2008}
\usepackage{graphicx}

\title{Multi-line  Analysis of Stellar Spectra in the VO Environment}
\author{Petr \v{S}koda}
\affil{Astronomical Institute of the Academy of Sciences of the Czech
Republic, v.v.i. \\ Fri\v{c}ova 298, Ond\v{r}ejov, CZ-251\,65}

\begin{document}

\maketitle
\begin{abstract} 
Despite the lack of important functions supported only by legacy applications,
the current VO-compatible tools have enough capabilities to allow powerful
analysis of stellar spectra using both public archives and local proprietary
data. We give examples of the possible multi-wavelength analysis of changes of
profiles in  different spectral lines  using the current SSAP-compatible VO
tools. The potential of future VO applications supporting more elaborate
methods for stellar analysis is discussed as well. 
\end{abstract}
\section{Introduction} 

Astronomical spectroscopy uses a wide range of techniques with different level
of complexity to achieve its final goal --- to  estimate the most precise and
reliable  information about celestial objects. 

The  scientific analysis of such  spectra requires further processing by the
variety of different methods. In certain studies a huge number of spectra has
to be collected from different servers (in different spectral regions)
and transformed into  common units. 

Accomplishing the multi-spectral analysis in VO environment may benefit from
automatic aggregation of distributed archive resources, seamless on-the-fly
data conversion, common interoperability of all tools  (using PLASTIC or SAMP
protocol) and powerful graphical visualisation of measured and derived
quantities
\section{Information obtained from spectroscopic analysis}
The spectral lines give us a plenty of information about the physical nature
and evolution of the given astronomical object. Examples of information
estimated from different parameters of spectral lines are given below:
\begin{itemize}
\item {\bf Position of line (wavelength)}
\begin{itemize}
\item Individual chemical elements present in the stellar atmosphere
\item Excitation / ionisation state of given element
\item Structure of molecules (e.g. rotational and vibrational states of molecules in IR regions)
\item Radial velocity (RV). If variable RV is measured, it may be the sign of binarity, sometimes even the  estimation of orbital parameters is feasible
\end{itemize}
\item {\bf Line Shape}
\begin{itemize}
\item Stellar parameters ($T_{\rm eff}$, $\log\,g$, rotation)
\item Stellar activity (turbulence, granulation)
\item The profiles of core/wings can present different physics dependent on optical depth (e.g. limb darkening law)
\item Expansion of gas shells, winds (P Cyg profiles in Novae, double-peaked emission profiles in Be stars)
\end{itemize}
\item {\bf Line profile variability in time} 
\begin{itemize} 
\item Change of physical state (e.g trigger of  of emission phase in Be stars,
outbursts of novae) 
\item Stellar spots (magnetic field structures, local overabundance  - typical for Ap stars) 
\item Pulsations ($\delta$\ Cep, RR Lyr, Miras)
\item Non radial pulsations (NRP) (e.g. $\delta$\ Sct, $\beta$\ Cep types)
\item Detection of extra-solar planets in spectra (bisector method makes small contributions
in global line profile enlarged and variable during planet's transit) 
\end{itemize} 
\end{itemize}
\section{Spectra post-processing}
The current VO applications using SSA protocol are very simple providing only
the URL of the given dataset, so the client gets a whole (often quite large)
file. The further processing is then done fully by the client which has to
download all the relevant datasets and keep them in the memory or local
storage.  The real strength of the VO technology lies, however, in transferring
part of the client's work to the server, which usually runs on a powerful
machine with fast connection to the data archives.  The typical examples of
commonly required post-processing of fully reduced spectra (at least in stellar
astronomy) are given below:
\begin{itemize} 
\item Cutout services (selection of only  certain spectral lines or regions
within the given wavelength range)  
\item Projection of multidimensional datasets (in 3D spectroscopy) 
\item Rectification (normalisation) of continuum 
\item Rebinning to given, usually equidistant grid of wavelengths (constant $\Delta\lambda$ or $\Delta\ln\lambda$) 
\item (De)convolution of instrumental  profile 
\item Application of physical broadening functions (rotation, limb darkening)
\item Shift in radial velocity, application of heliocentric correction computed on server 
\item Merging of individual echelle orders in a single spectrum
\end{itemize}
Although most of these post-processing methods can be implemented in a
straightforward way, the rectification of continua may be a difficult problem, 
especially on echelle spectra.
\section{Advanced spectral analysis}
As the goal of the VO is to make easy and comfortable the physical analysis of a
huge number of fully reduced (and post-processed) spectra in the environment of
VO client (or web portal), the common recipes of spectral analysis have to be
implemented as VO-compatible.  The basic techniques were reviewed in
\citet{Sko2008}, we will briefly mention only the more complicated 
tasks of spectra analysis: 
\subsection{Dynamical  spectrum } 
It is sometimes called the gray representation or trailed spectrum. The basic
idea is to find the small time-dependent deviations  of individual line
profiles from an averaged one. 
\begin{figure} 
\centering 
\includegraphics[width=1.00\linewidth]{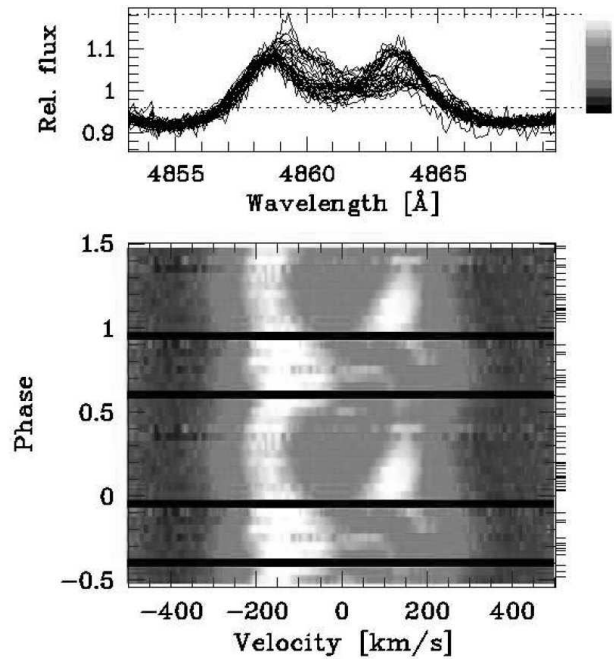}
\caption{Dynamical spectrum of H$\beta$  line profile variability of 59~Cyg.
Residuals from average profile of 38 spectra  are binned to 20 phase bins
corresponding to period 28.192 days. Two expanded cycles shown for clarity.
Individual profiles are overplotted above.  After \citet{maintz:2003}
\label{dynspec2}}
\end{figure}
First the average of many high dispersion high SNR spectra (with removal of
outliers) is prepared (called template spectrum). Then each individual spectrum
in time series is either divided by the template (quotient spectrum) or the
template is subtracted from it (the differential spectrum). The group of
similar resulting intensities is given the same colour or  level of gray.  See
Fig.~\ref{dynspec2}.\hspace{0.5em} More examples may be found e.g.  in
\citet{1999A&A...345..172D}, \citet{maintz:2003} or \cite{uytterhoeven:2004}.
\subsection{Measurement of radial velocity and higher moments of line profile}
The one of the important information received from spectrum is the radial
velocity (RV) of the object. From its changes the binarity can be revealed, or
the possession of extrasolar planet.  The combination of higher moments of line
profile is a one of the possible ways of determination of non radial pulsation
modes -- numbers $l,m$ \citep{1992A&A...266..294A}. 
\subsection{Measurement of equivalent width}
The Equivalent width (EW) of the spectral line gives the information about the
number of absorbing or emitting atoms of given element.  Emission lines by
definition have negative EW.  The changes in EW during time may bring about the
information about the dynamic evolution of the target and may be subjected to
period analysis.  
\subsection{Bisector analysis}
It is a method describing quantitatively the tiny asymmetry or subtle changes
in line profiles.  The characteristic shape of  bisector gives the information
about turbulence fields (e.g. convection) in stellar photosphere, characterised
by the value of micro-turbulent velocity \citep{1982ApJ...255..200G,
2005PASP..117..711G} or about other processes causing the tiny profile
asymmetry.  It has been used successfully for searching of extrasolar planets
\citep{2001AJ....121.1136P} or in asteroseismology.  It requires high resolution
normalised spectra with extremely high SNR. 
\subsection{Period Analysis}
Its aim is to find the hidden periods of variability of given object.
Sometimes this period can be identified with some physical mechanism (e.g.
orbital period of binaries, rotational modulation or pulsations). Wide range of
objects show the multi-periodicity on various time scales (e.g. binary with
pulsating components).  Ones the suspected period is found, the data may be
folded accordingly, plotted in circular phase corresponding to this period. 
\subsection{Doppler imaging }
It was introduced by \citet{1983PASP...95..565V} as a method allowing the
surface mapping of stellar spots.  First test were done on stars of RS CVn type
and on $\zeta$~Oph \citep{1983ApJ...275..661V}. Works well on rapid rotators
and needs a high resolution spectra with very high SNR (300--500). Tho whole
rotational period should be covered well, better several times.  When all the
requirements are met,  the map  of surface features (spots, nodes of non radial
pulsations) is obtained with very high accuracy. See the left panel of
Fig.~\ref{zeemandop}.
\subsection{Zeeman Doppler Imaging}
\begin{figure} 
\centering 
\includegraphics[width=1.00\linewidth]{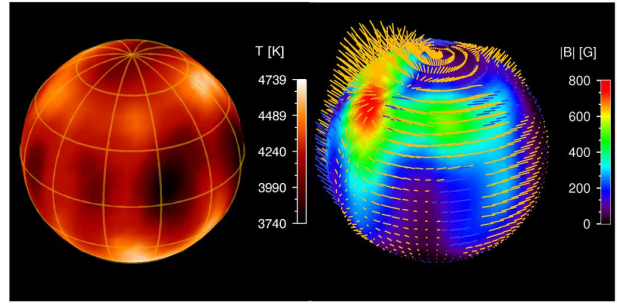}
\caption{Zeeman Doppler Imaging of II Peg.
After \citet{Strassmeier2008}
\label{zeemandop}}
\end{figure}

Quite complicated processing of spectra is required for study of stellar
magnetic fields.  The estimation of magnetic field from  polarimetry using the
Zeeman phenomenon involves the processing of long series of homogeneous spectra
to be accomplished in parallel with extreme precision and requires again the
information from synthetic models (simulation of Stokes parameters on simulated
magnetic stars) The nice example is the model of II Peg by
\citet{Strassmeier2008}. See the right panel of Fig.~\ref{zeemandop}.
\subsection{Spectra disentangling}
This method allows to separate the spectra of individual stars in binary or
multiple systems and simultaneously to find orbital parameters of the system,
even in case of heavy blending of lines.  It supposes the changes in line
profile are caused only by combination of Doppler shifted components (no
intrinsic variability of star).  The best solution of orbital parameters and
disentangled line profiles of individual stellar components are found by least
square global minimisation.  The method also enables the removal of the telluric
lines with great precision.  The good orbital coverage and the estimate of
orbital parameters is required. Two approaches exist:  

\noindent {\it Wavelength space disentangling}\\
developed by \citet{1991ApJ...376..266B} and improved by
\citet{1994A&A...281..286S}.  It needs a large memory to store sparse matrices,
requires large computing power.  It is more straightforward to understand the
results and sources of errors.

\noindent {\it Fourier space disentangling} \\
introduced by \citet{1995A&AS..114..393H,1997A&AS..122..581H} in program KOREL.
Another program available today (still based on KOREL ideas) is FDBINARY
\citep{2004ASPC..318..111I}.  They  work in Fourier space, and transform the
wavelengths into $\ln\,\lambda$.  They solve a small amount of linear
equations, so they are   memory savvy and can be run on even small computer.
The method, however,  requires  perfect continuum fit.
\subsection{Classification of stellar spectra}
It used to be a very time-consuming and quite subjective method in the era of
photographic plates (e.g. MKK classification).  Today it has been done finding
the minimal differences between a grid of template (or even synthetic) spectra
and the examined one.  The techniques of artificial intelligence (e.g. neural
networks) may be very efficient, but the global optimisation using the $\chi^2$
minimisation seems to be more reliable with possibility of iterative control.
An example of automatic classification engine for white dwarfs is  described by
\citet{Winter2006}.
\section{Killer VO spectral applications}

Here we give some scientific cases that may become the killer application
forcing the  wide astronomical community  to use the VO tools with capabilities
suggested above despite its current reluctance and hesitation:
\begin{itemize} 
\item Use VO to find all stars with emission in given line, find the time when
it was in emission and plot the time evolution of its EW.
\item Use VO to get 1000 or more spectra of the given object, cut out regions
around given lines, plot the lines, make a gray dynamic spectrum folded in time
\item The same, but fold by period clicked on interactive periodogram
\item Get the unknown lines identification of piece of spectra from theoretical
observatory (SLAP - Simple line access protocol) having the line selection
limited by pre-estimated temperature (using Saha equation)
\item Create light and radial velocity curve of a binary star for given period
(estimated by other VO tool running in parallel and exchanging data over
PLASTIC or SAMP)
\item Fit the grid of models ($T_{\rm eff}, \log\, g$) to the observed spectrum
for many stars. 
\item Extract detailed profiles of given spectral lines from archives of
echelle spectra where every echelle order is kept separately (unmerged spectra
in so called pixel-order space)
\end{itemize}
\section{Conclusions}
The  large part of spectroscopic analysis today  has been accomplished by
several independent non VO-compatible legacy packages as the currently
available VO clients supporting the SSA protocol can provide with only simple,
mostly interactive capabilities on individual spectra.

The processing of large amount of spectra in a consistent way requires the
simple post-processing (e.g. cutout or rebinning) services to be implemented on
the server side

A lot of interesting physical information about astronomical objected can be
obtained from only several short spectral ranges covering selected spectral
lines. Dynamic behaviour of interesting objects can be estimated from the
changes of line profiles as well as  from period analysis of their moments
(including equivalent width). 

By introduction of  modern  VO-aware tools into the  astronomical spectral
analysis  a remarkable increase of effectiveness of astronomical research can
be achieved.
\section*{Acknowledgements}
This work has been supported by grant GACR 205/06/0584 and EURO-VO DCA WP6 and AIDA funds.
The Astronomical Institute Ond\v{r}ejov is supported by project AV0Z10030501
\bibliographystyle{aa}

\end{document}